\documentstyle[preprint,eqsecnum,aps,prb]{revtex}
\begin{document}
\draft
\title{\Large\bf PERFECT HYPERMOMENTUM FLUID: VARIATIONAL THEORY \\
                 AND EQUATIONS OF MOTION}
\author{Olga V Babourova\dag\footnote[3]{babourova@orc.ru} and
Boris N Frolov\ddag\footnote[4]{frolovbn@orc.ru}}

\address{\dag\ Department of Theoretical Physics, Faculty of Physics,
Moscow State University, Leninskie Gory, d.1, st.2, Moscow 119992, Russian Federation}

\address{\ddag\ Department of Physics, Faculty of Mathematics, Moscow State
Pedagogical University, Krasnoprudnaya 14, Moscow 107140, Russian Federation}
\maketitle
\begin{abstract}
     The variational theory of the perfect hypermomentum fluid is
developed. The new type of the generalized Frenkel condition is considered.
The Lagran\-gi\-an density of such fluid is stated, and the equations of
motion of the fluid and the Weyssenhoff-type evolution equation of the
hypermomentum tensor are derived. The expressions of the matter currents of
the fluid (the canonical energy-momentum 3-form, the metric stress-energy
4-form and the hypermomentum 3-form) are obtained. The Euler-type
hydrodynamic equation of motion of the perfect hypermomentum fluid is derived.
 It is proved that the motion of the perfect fluid without hypermomentum in a
metric-affine space coincides with the motion of this fluid in a Riemann
space.
\end{abstract}
\newpage

\section{Introduction}
\markright{Perfect hypermomentum fluid}
\setcounter{equation}{0}
     The Weyssenhoff--Raabe perfect spin fluid\cite{WR,Hal} has a wide range
of application in modern cosmology and astrophysics.\cite{Tr}${}^{-}$\cite{Gas}
The variational theory of this fluid is based on accounting the constraints in
the Lagrangian density of the fluid with the help of Lagrange multipliers.
Such theory was developed in case of a Riemann--Cartan space in Refs.
\onlinecite{Tun1} --\onlinecite{Ol} and in case of a metric-affine space in
Refs.\onlinecite{Bab-Fr},\onlinecite{Bab:thes}. On the other variational
methods of the perfect spin fluid in a Riemann--Cartan space see Refs.
\onlinecite {Min-Kar} and \onlinecite{Kop}.
\par
     The Weyssenhoff--Raabe  perfect  spin  fluid  represents the particular
case of the theory of the perfect fluid with intrinsic degrees of freedom.
The natural  generalization  of the Weyssenhoff--Raabe perfect spin fluid is
the perfect spin fluid with color charge, every particle of which is endowed
with spin and intrinsic non-Abelian color charge.\cite{GHK}${}^{-}$\cite{Yaf}
The variational theory of the fluid of such type was constructed in Refs.
\onlinecite{Am} --\onlinecite{Ind},\onlinecite{Bab:thes},\onlinecite{DAN},
\onlinecite{Yaf}. This model of the fluid can be used in nuclear physics for
the generalization of the hydrodynamical description of adron multiple
appearance\cite{VP} (Landau model) and also in quark-gluon plasma physics and
astrophysics.
\par
     The other significant generalization of the Weyssenhoff--Raabe  perfect
spin fluid  is  the  perfect dilaton-spin fluid,  every particle of which is
endowed with intrinsic spin and dilatonic charge.\cite{dil}${}^{,}$\cite{part}
Significance of matter with dilatonic charge is based on  the  fact  that  a
low-energy effective  string  theory is reduced to the theory of interacting
metric and dilatonic field.\cite{GSW}
\par
     The essential  meaning  of the fluid models discussed above consists in
the fact that dynamics of this type of matter imposes the constraints  on  a
metric  and  a  connection  of  the  space-time  manifold  and  generates the
Riemann--Cartan or the Weyl--Cartan  geometrical  structures  of  space-time.
\par
     The generalization of the spin and dilaton dynamical variables is the
hypermomentum tensor $J^{\alpha}\!_{\beta}$ introduced in
Ref.\onlinecite{He-Ker}. It has the decomposition,
\begin{eqnarray}
&&J_{\alpha\beta} = S_{\alpha\beta} + \hat J_{(\alpha\beta)} + \frac{1}{4}
g_{\alpha\beta}J\;, \label{eq:hyp} \\ &&S_{\alpha\beta}:= J_{[\alpha\beta]}\;,
\qquad J := g^{\mu\nu}J_{\mu\nu}\;,\qquad \hat J_{(\alpha\beta)} =
J_{(\alpha\beta)} - \frac{1}{4}g_{\alpha\beta}J\;,\label{eq:SJ}
\end{eqnarray}
where $S_{\alpha\beta}$ is the spin tensor,  $J$ is the dilatonic charge and
$\hat  J_{(\alpha\beta)}$  is  the  intrinsic  proper  hypermomentum (shear)
tensor.
\par
     The natural generalization of the perfect spin fluid and the perfect
dilaton-spin fluid is the perfect fluid,  every particle of which is endowed
with intrinsic hypermomentum. This new hypothetical type of matter was
announced in  Refs.\onlinecite{Bab1},\onlinecite{Bab:thes} and named
{\it perfect fluid with intrinsic hypermomentum}. The variational theory of
such fluid was developed by various
authors.\cite{Bab:Arg}${}^{-}$\cite{Los:pr}
In Refs.\onlinecite{Ob-Tr},\onlinecite{He:pr} this type of matter was named
{\it hyperfluid}. We shall name this type of fluid as {\it perfect
hypermomentum fluid}.
\par
  The real existence of the hypermomentum fluid is the fact  of  fundumental
meaning because as a source of gravitational field it generates the new type
of  space-time  geometry,  namely  the  geometry  of  a  metric-affine  space
$(L_{4},g)$  (see Ref.\onlinecite{He:pr} and references therein).  Nowadays,
the metric-affine theory of gravitation arouses an  interest  in  connection
with  the  problem  of  the  relation  of the gravitation and the elementary
particles physics\cite{Ne:Si} and with the problem of  renormalizability  of
the gravitational theory.\cite{LN}
\par
     We shall  develop  the   phenomenological   macroscopic   approach   in
the description of the perfect hypermomentum  fluid.  In  this  approach the
physical variables which characterize the fluid  (such  as  energy  density,
pressure, entropy, hypermomentum ets.) are considered to be smooth functions
of space-time coordinates.  This approach is compatible with the  statistical
continuous  medium  approach by means of the assumption that a fluid element
represents a  statistical  subsystem  with  sufficiently  large  number   of
particles.  The fluid element can be considered as a quasiclosed system with
the properties that coincide with macroscopic properties of the  fluid.  The
physical  variables characterizing the fluid are regarded to be averaged over
the fluid element.
\par
     The theory of the perfect fluid with intrinsic degrees of freedom being
developed,  the additional intrinsic degrees of freedom of a  fluid  element
are described by the four vectors $\bar{l}_{p}$ ($p = 1,2,3,4$), called {\it
directors},  attached with each element  of  the  fluid.  Three  of  the
directors  ($p  =  1,2,3$)  are  space-like  and the fourth one ($p = 4$) is
time-like.
\par
     In Riemann and  Riemann--Cartan  spaces  a  fluid  element endowed with
directors  moves  according  to  the  Fermi  transport  that  preserves  the
orthonormalization  of the directors.  In a metric-affine space,  in which a
metric and a connection are not compatible  it  is  naturally  consider  the
directors to be {\it elastic}\cite{Ob-Tr,Bab:tr3,Sm-Kr} in the sense that they
can  undergo  arbitrary  deformations  during  the  motion  of  the   fluid.
Nevertheless, in most theories it is accepted that the time-like director is
collinear to the 4-velocity  $u^{\alpha}$  of  the  fluid  element and
orthogonality of space-like directors to a 4-velocity is maintained.
\par
     The distinction of the variational machinery consists in using the
generalized Frenkel condition,\cite{Bab1}${}^{,}$\cite{Bab:Arg}${}^{,}$
\cite{Ob-Tr}
\begin{equation}      \label{eq:01}
J_{\alpha\beta} u^{\beta} = 0\; , \qquad J_{\alpha\beta}u^{\alpha} = 0 \; ,
\end{equation}
or the Frenkel condition in its standard classical
form,\cite{Bab:tr3,GR14,BF:Los,Nov}
\begin{equation}   \label{eq:02}
S_{\alpha\beta} u^{\beta} = 0 \; ,
\end{equation}
where $J_{\alpha\beta}$  and  $S_{\alpha\beta}$  are   the   specific   (per
particle)  intrinsic  hypermomentum tensor and the specific spin tensor of a
fluid   element,   respectively.   Another   possibility   is   so    called
``unconstrained   hyperfluid''\cite{Ob2},  in  which  any  type  of  Frenkel
condition is absent.
\par
     In Ref.\onlinecite{Ob2} it is mentioned that in case of the generelized
Frenkel  condition  (\ref{eq:01}) the dilatonic charge of a fluid element is
expressed in  terms   of   the   shear   tensor   (see   the   decomposition
(\ref{eq:hyp})): $J  =  (4/c^2)\hat J_{\alpha\beta}u^{\alpha}u^{\beta}$. In
this case the hypermomentum fluid can not be of  the  pure  dilationic  type
with $\hat  J_{\alpha\beta}  = 0$ and $J \not = 0$.  On the other hand,  the
Frenkel condition (\ref{eq:02}) leads to the unusual form of  the  evolution
equation of  the  hypermomentum  tensor,\cite{GR14,BF:Los}  which  does  not
demonstrate the  Weyssenhoff-type  dynamics.   As   to   the   unconstrained
hyperfluid, in  Ref.~\onlinecite{Ob2}  it  is stated that such type of fluid
does not contain the Weyssenhoff spin fluid as a particular case.  Therefore
all three  kinds  of  the  approaches  mentioned  are  not satisfactory from
physical point of view.
\par
     In this paper we consider the  new  type  of  the  generalized  Frenkel
condition,  which allows to construct the hypermomentum perfect fluid theory
with the dilaton-spin fluid and the Weyssenhoff  spin  fluid  as  particular
cases.  In  our  approach  it  is also essential that all four directors are
elastic.  None of the orthogonality conditions  of  the  four  directors  is
maintained during the motion of the fluid.  Besides,  the time-like director
needs not to be collinear to the 4-velocity of the  fluid  element.  On  the
preliminary version of our results see Ref.\onlinecite{Los:pr}.
\par
     Our paper is organized as follows.
\par
     In Sec.  2  the  dynamical  variables and constraints of the theory are
discussed. In Sec. 3 the Lagrangian density of the perfect hypermomentum
fluid is  stated,  and the equations of motion of the fluid are derived.  In
Sec. 4 the Weyssenhoff-type evolution equation of the hypermomentum tensor
is stated. Then in Sec. 5 the expressions of the matter currents of the
hypermomentum fluid (the canonical energy-momentum 3-form, the metric
stress-energy 4-form and  the  hypermomentum 3-form) are obtained.  Sec. 6 is
devoted to the derivation of the Euler-type hydrodynamic equation of motion of
the perfect hypermomentum fluid. At last the peculiarities of the
hypermomentum fluid motion are discussed in Sec. 7.
\par
     We use  the  exterior  form  variational  method  according to Trautman
\cite{Tr1,Tr2}.

\section{The dynamical variables and constraints}
\markright{Perfect hypermomentum fluid}
\setcounter{equation}{0}
     In the exterior form language the material frame of the directors
turns into the coframe of 1-forms $l^{p}$ $(p = 1,2,3,4)$, which have dual
3-forms $l_{q}$, while the constraint
\begin{equation}
l^{p} \wedge l_{q} = \delta^{p}_{q}\eta\;, \qquad
l^{p}_{\alpha}l_{p}^{\beta} = \delta_{\alpha}^{\beta}\;, \label{eq:1}
\end{equation}
being fulfilled, where $\eta$ is the volume 4-form and the component
representations are introduced,
\begin{equation}
l^{p} = l^{p}_{\alpha}\theta^{\alpha}\; , \;\;\;\;\;
l_{q} = l_{q}^{\beta}\eta_{\beta}\; .\label{eq:2}
\end{equation}
Here $\theta^{\alpha}$ is a 1-form basis and $\eta_{\beta}$ is a 3-form
defined as \cite{Tr1}
\begin{equation}
\eta_{\beta} = \bar{e}_{\beta}\rfloor \eta = *\theta_{\beta} \; ,
\;\;\;\;\;\;\;\;\;\; \theta^{\alpha} \wedge \eta_{\beta} = \delta^{\alpha}
_{\beta} \eta \; , \label{eq:3}
\end{equation}
where $\rfloor$ means the interior product, $*$ is the Hodge dual operator
and $\bar{e}_{\beta}$ is a basis vector, a coordinate system being
nonholonomic in general.
\par
     Each fluid element possesses a 4-velocity vector $\bar{u}=u^{\alpha}
\bar{e}_{\alpha}$ which is corresponded to a flow 3-form $u$ (Ref.
\onlinecite{Tr2}), $u:=\bar{u}\rfloor\eta=u^{\alpha}\eta_{\alpha}$ and a
velocity 1-form $*\! u=u_{\alpha}\theta^{\alpha}=g(\bar{u},\underline{})$
with
\begin{equation}
*\!u \wedge u = -c^{2}\eta \;, \label{eq:4}
\end{equation}
that means the usual condition $g(\bar{u},\bar{u}) = - c^{2}$, where
$g(\underline{}, \underline{})$ is the metric tensor.
\par
     A fluid element moving, the fluid particles number and entropy
conservation laws are fulfilled,
\begin{eqnarray}
d(n u) = 0 \; , \qquad d(n s u) = 0 \; , \label{eq:7}
\end{eqnarray}
where $n$ is the fluid particles concentration equal to the number of fluid
particles per a volume unit, and $s$ is the the specific (per particle)
entropy of the fluid in the rest frame of reference, respectively.
\par
     The measure of ability of a fluid element to perform the intrinsic
motion is the quantity $\Omega^{q}\!_{p}$ which generalizes the fluid element
``angular velocity'' of the Weyssenhoff spin fluid theory. It has the form
\begin{equation}
\Omega^{q}\!_{p} \eta := u \wedge l^{q}_{\alpha}
{\cal D} l^{\alpha}_{p} \; , \label{eq:9}
\end{equation}
where ${\cal D}$ is the exterior covariant differential with respect to a
connection 1-form $\Gamma^{\alpha}\!_{\beta}$,
\begin{equation}
{\cal D} l^{\alpha}_{p} = d l^{\alpha}_{p} + \Gamma^{\alpha}\!_{\beta}
l^{\beta}_{p} \; . \label{eq:10}
\end{equation}
\par
     An element of the fluid with intrinsic hypermomentum possesses the
additional ``kinetic'' energy 4-form,
\begin{equation}
E =  \frac{1}{2} n J^{p}\!_{q}\Omega^{q}\!_{p}\eta = \frac{1}{2}n
J^{p}\!_{q} u\wedge l^{q}_{\alpha}{\cal D} l^{\alpha}_{p} \; , \label{eq:8}
\end{equation}
where $J^{p}\!_{q}:=J^{\alpha}\!_{\beta} l^{p}_{\alpha} l^{\beta}_{q}$ is the
specific intrinsic hypermomentum tensor representing the new dynamical
quantity which generalizes the spin density of the Weyssenhoff fluid.
\par
     The hypermomentum tensor $J^{p}\!_{q}$ can be decomposed into
irreducible parts,
\begin{eqnarray}
&&J^{p}\!_{q} = \hat J^{p}\!_{q} + \frac{1}{4} \delta^{p}_{q}J\;, \qquad
J := J^{p}\!_{p}\; ,\qquad \hat J^{p}\!_{p} = 0\; ,  \label{eq:11} \\
&& \hat J^{p}\!_{q} := S^{p}\!_{q} + \hat J^{(p}\!_{q)}\; , \qquad
S^{p}\!_{q}:= J^{[p}\!_{q]}\;, \qquad \hat J^{(p}\!_{q)} = J^{(p}\!_{q)}
- \frac{1}{4} \delta^{p}_{q}J\; . \label{eq:112}
\end{eqnarray}
Here $S^{p}\!_{q}$ is the specific spin tensor, $J$ is the specific dilatonic
charge and $\hat J^{(p}\!_{q)}$ is the specific intrinsic proper
hypermomentum (shear) tensor of a fluid element, respectively. We shall name
the quantity $\hat J^{p}\!_{q}$ as the specific {\it traceless hypermomentum
tensor}.
\par
     It is well-known that the spin tensor is spacelike in its nature that
is the fact of fundamental physical meaning. This leads to the classical
Frenkel condition, $S^{\alpha}\!_{\beta} u^{\beta} = 0$. We shall suppose
here that the full traceless part of the hypermomentum tensor
$\hat J^{p}\!_{q}$ (not only the spin tensor but also the tensor
$\hat J^{(p}\!_{q)}$) has such property and therefore satisfies the
generalized Frenkel conditions in the form,
\begin{eqnarray}
&\hat J^{p}\!_{q}u_{p} = 0\; , \qquad u_{p}:= u_{\alpha} l^{ \alpha}_{p}\;,
\label{eq:110}\\
&\hat J^{p}\!_{q} u^{q} = 0\; , \qquad u^{q} := u^{\alpha} l_{\alpha}^{q}\;,
\label{eq:111}
\end{eqnarray}
which can be written in the following way,
\begin{eqnarray}
&\hat J^{p}\!_{q} l_{p}\wedge *\!u   = 0 \; , \label{eq:12} \\
&\hat J^{p}\!_{q} l^{q}\wedge u = 0 \; . \label{eq:120}
\end{eqnarray}
The Frenkel conditions (\ref{eq:110}), (\ref{eq:111}) are equivalent to
the equality,
\begin{equation}
\Pi^{p}_{r} \Pi^{t}_{q} \hat J^{r}\!_{t} = \hat J^{p}\!_{q}\;, \quad
\Pi^{p}_{r} := \delta^{p}_{r} + \frac{1}{c^{2}} u^{p} u_{r}\;. \label{eq:121}
\end{equation}
Here $\Pi^{p}_{r}$ is the projection tensor, which separates the subspace
orthogonal to the fluid velocity.
\par
     The internal energy density of the fluid $\varepsilon$ depends on
the extensive (additive) ther\-mo\-dy\-namic parameters $n$, $s$, $J^{p}\!_
{q}$ and obeys to the first thermodynamic principle,
\begin{equation}
d\varepsilon(n, s, J^{p}\!_{q}) = \frac{\varepsilon + p}{n} dn +
n T ds + \frac{\partial \varepsilon}{\partial J^{p}\!_{q}} dJ^{p}\!_{q}
\; , \label{eq:13}
\end{equation}
where $p$ is the hydrodynamic fluid pressure and $T$ is the temperature.
\par
      We shall consider as independent variables the quantities $n$, $s$,
$J^{p}\!_{q}$, $u$, $l^{q}$, $\theta^{\sigma}$, $\Gamma^{\beta}\!_{\alpha}$,
the constraints (\ref{eq:4}), (\ref{eq:7}), (\ref{eq:12}), (\ref{eq:120})
being taken into account in the Lagrangian density by means of the Lagrange
multipliers.
\par
     In what follows we need the variation,
\begin{equation}
\delta \eta = \eta \frac{1}{2} g^{\alpha\beta}\delta g_{\alpha\beta} +
\delta \theta^{\sigma} \wedge \eta_{\sigma} \;. \label{eq:17}
\end{equation}
As a result of the relation $\theta^{\alpha}\wedge u = u^{\alpha}\eta $
one has,
\begin{equation}
\eta \delta u^{\alpha} = - \delta u \wedge \theta^{\alpha} + \delta \theta^
{\alpha} \wedge u - u^{\alpha}\delta \eta\; . \label{eq:16}
\end{equation}
The relation $*\! u = g_{\alpha\beta}u^{\alpha}\theta^{\beta}$ yields the
variation,
\begin{equation}
\delta *\! u = g_{\alpha\beta}\theta^{\alpha}\delta u^{\beta} +
u^{\alpha}\theta^{\beta}\delta g_{\alpha\beta} + u^{\beta}g_{\sigma\beta}
\delta \theta^{\sigma}\; . \label{eq:161}
\end{equation}
As a result of the resolution of the constraints (\ref{eq:1}) and with the
help of the relations (\ref{eq:3}), one can derive the variations,
\begin{eqnarray}
&&\eta \delta l^{p}_{\alpha} = - \delta \theta^{\sigma}\wedge \eta_{\alpha}
l^{p}_{\sigma} + \delta l^{p} \wedge \eta_{\alpha}\;, \label{eq:14} \\
&&\eta \delta l_{p}^{\alpha} = \delta \theta^{\alpha} \wedge l_{p}
- \delta l^q \wedge l_{q}^{\alpha} l_{p}\; . \label{eq:15}
\end{eqnarray}

\section{The Lagrangian density and the equations of motion of the
     fluid}
\markright{Perfect hypermomentum fluid}
\setcounter{equation}{0}
     The perfect fluid Lagrangian density 4-form of the perfect
hypermomentum fluid should be chosen as the remainder after subtraction the
internal energy density of the fluid $\varepsilon$ from the ``kinetic'' energy
(\ref{eq:8}) with regard to the constraints (\ref{eq:4}), (\ref{eq:7}),
(\ref{eq:12}), (\ref{eq:120}) which should be introduced into the Lagrangian
density by means of the Lagrange multipliers $\lambda$, $\varphi$, $\tau$,
$\chi^{q}$, $\zeta_{p}$, respectively. As a result of the previous section
the Lagrangian density 4-form has the form
\begin{eqnarray}
{\cal L}_{m} = L_{m} \eta = - \varepsilon (n, s, J^{p}\!_{q}) \eta +
\frac{1}{2}n J^{p}\!_{q} u\wedge l^{q}_{\alpha}{\cal D} l^{\alpha}_{p}
+ n u\wedge d\varphi + n\tau u \wedge ds \nonumber\\
+ n \lambda (*\!u \wedge u + c^2 \eta) + n \chi^q\hat J^p\!_q l_p\wedge *\!u
+ n\zeta_p \hat J^p\!_q l^q\wedge u\;. \label{eq:20}
\end{eqnarray}
\par
     The fluid motion equations and the evolution equation of  the
hypermomentum tensor are derived by the variation of (\ref{eq:20}) with
respect to the independent variables $n$, $s$, $J^{p}\!_{q}$, $u$,
$l^{q}$, and the Lagrange multipliers, the thermodynamic principle
(\ref{eq:13}) being taken into account.  We shall consider the 1-form  $l^q$
as an  independent  variable  and the 3-form $l_p$ as a function of $l^q$ by
means of (\ref{eq:1}). As a result of such variational machinery one gets the
constraints (\ref{eq:4}), (\ref{eq:7}), (\ref{eq:12}), (\ref{eq:120}) and the
following variational equations,
\begin{eqnarray}
\delta n : &&\quad (\varepsilon + p) \eta - \frac{1}{2}n
J^{p}\!_{q} u\wedge l^{q}_{\alpha}{\cal D} l^{\alpha}_{p}
- n u\wedge d\varphi = 0 \; ,\label{eq:21}\\
\delta s : &&\quad  T \eta + u \wedge d\tau = 0 \;, \label{eq:22}\\
\delta J^{p}\!_{q} : &&\quad \frac{\partial \varepsilon}{\partial J^{p}\!_{q}}
= \frac{1}{2}n\Omega^{q}\!_{p} - n(\chi^{q}u_{p} - \zeta_{p}u^{q}) +
\frac{1}{4}n\delta^{p}_{q} (\chi^{r}u_{r} - \zeta_{r}u^{r})\;,\label{eq:23}\\
\delta u : &&\quad d\varphi + \tau ds - 2\lambda *\! u + \chi^{q}\hat
J_{\beta q}\theta^{\beta} - \zeta_{p}\hat J^{p}\!_{q} l^{q}
+ \frac{1}{2}J^{p}\!_{q} l^{q}_{\alpha} {\cal D}l^{\alpha}_{p} = 0\; ,
\label{eq:25} \\
\delta l^{q} : &&\quad \frac{1}{2}\dot{J}^{\sigma}\!_{\rho}l_{q}^{\rho}
\eta_{\sigma} - \chi^{r}\hat J^{p}\!_{r} u_q l_p
 - \zeta_{r} \hat J^{r}\!_{q} u = 0\;. \label{eq:26}
\end{eqnarray}
Here the ``dot'' notation for the tensor object $\Phi$ is introduced,
\begin{equation}
\dot{\Phi}^{\alpha}\!_{\beta} := *\!(u\wedge {\cal D}\Phi^{\alpha}\!_
{\beta})\; . \label{eq:27}
\end{equation}
\par
     Multiplying the equation (\ref{eq:25}) by $u$ from the left externally
and using (\ref{eq:7}) and (\ref{eq:21}), one derives the expression for the
Lagrange multiplier $\lambda$,
\begin{equation}
2 n c^{2} \lambda = \varepsilon + p \; . \label{eq:28}
\end{equation}
\par
     As a consequence of the equation (\ref{eq:21}) and the constraints
(\ref{eq:4}), (\ref{eq:7}), (\ref{eq:12}), (\ref{eq:120}) one can verify that
the Lagrangian density 4-form (\ref{eq:20}) is proportional to the
hydrodynamic fluid pressure, ${\cal L}_{m} = p \eta$, which corresponds to
Ref.\onlinecite{Rit}.

\section{The evolution equation of the hypermomentum tensor}
\markright{Perfect hypermomentum fluid}
\setcounter{equation}{0}
     The variational equation (\ref{eq:26}) represents the evolution equation
of the hypermomentum tensor. Multiplying the equation (\ref{eq:26}) by
$l^{p}_{\beta}\theta^{\alpha}\wedge\dots$ from the left externally one gets,
\begin{equation}
\frac{1}{2}\dot{J}^{\alpha}\!_{\beta} - \chi^{r}\hat J^{\alpha}\!_{r}
u_{\beta} - \zeta_{r}\hat J^{r}\!_{\beta}u^{\alpha} = 0\; . \label{eq:30}
\end{equation}
Contractions (\ref{eq:30}) with $u_{\alpha}$ and then with $u^{\beta}$ yield
the expressions for the Lagrange multipliers,
\begin{eqnarray}
&& \zeta_{r}\hat J^{r}\!_{\beta} = - \frac{1}{2c^2} \dot{J}^{\gamma}
\!_{\beta} u_{\gamma}\; , \label{eq:31} \\
&& \chi^{r}\hat J^{\alpha}\!_{r} = - \frac{1}{2c^2} \dot{J}^{\alpha}\!_
{\gamma}u^{\gamma}\; . \label{eq:32}
\end{eqnarray}
After the substitution of (\ref{eq:31}) and (\ref{eq:32}) into (\ref{eq:30})
one gets the evolution equation of the hypermomentum tensor,
\begin{equation}
\dot{J}^{\alpha}\!_{\beta} + \frac{1}{c^{2}} \dot {J}^{\alpha}\!_{\gamma}
u^{\gamma}u_\beta + \frac{1}{c^{2}} \dot {J}^{\gamma}\!_{\beta}
u_{\gamma}u^{\alpha} = 0\; . \label{eq:33}
\end{equation}
This equation generalizes the evolution equation of the spin tensor in the
Weyssenhoff fluid theory.
\par
     The equation (\ref{eq:33}) has the consequence,
\begin{equation}
\dot {J}^{\alpha}\!_{\beta} u_{\alpha} u^{\beta} = 0\; , \label{eq:34}
\end{equation}
which permits to represent the evolution equation of the hypermomentum tensor
(\ref{eq:33}) in the form,
\begin{equation}
\Pi^{\alpha}_{\sigma}\Pi^{\rho}_{\beta} \dot{J}^{\sigma}\!_{\rho} = 0 \;,
\label{eq:355}
\end{equation}
where the projection tensor $\Pi^{\alpha}_{\sigma}$ has been defined in
(\ref{eq:121}). The evolution equation of the hypermomentum tensor in the
form (\ref{eq:355}) was derived in Ref. \onlinecite{Nov}.
\par
     The contraction (\ref{eq:33}) on the indices $\alpha$ and $\beta$
gives with the help of (\ref{eq:34}) the dilatonic charge conservation law,
\begin{equation}
\dot{J} = 0\; . \label{eq:361}
\end{equation}

\section{The energy-momentum tensor of the perfect hypermomentum
     fluid}
\markright{Perfect hypermomentum fluid}
\setcounter{equation}{0}
     With the help of the matter Lagrangian density (\ref{eq:20}) one can
derive the external matter currents which are the sources of the
gravitational field. In case of the perfect hypermomentum fluid the matter
currents are the canonical energy-momentum 3-form $\Sigma_{\sigma}$,
the metric stress-energy 4-form $\sigma^{\alpha\beta}$  and the
hypermomentum 3-form ${\cal J}^{\alpha}\!_{\beta}$, which are determined as
variational derivatives.
\par
     The variational derivative of the explicit form of the Lagrangian
density (\ref{eq:20}) with respect to $\theta^{\sigma}$ yields the canonical
energy-momentum 3-form,
\begin{equation}
\Sigma_{\sigma} := \frac{\delta{\cal L}_{m}}{\delta \theta^{\sigma}}
 = - \varepsilon\eta_{\sigma} + 2\lambda n u_{\sigma} u
+ 2 c^{2}\lambda n \eta_{\sigma} - n \chi^{r}\hat J^{q}\!_{r}
(g_{\sigma\rho}l^{\rho}_{q} u + u_{\sigma}l_{q})
+ \frac{1}{2}n \dot{J}^{\rho}\!_{\sigma}\eta_{\rho}\; . \label{eq:350}
\end{equation}
Using the explicit form of the Lagrange multiplier (\ref{eq:28}), one gets,
\begin{equation}
\Sigma_{\sigma} =  p \eta_{\sigma} + \frac{1}{c^{2}}(\varepsilon + p)
u_{\sigma} u + \frac{1}{2}n \dot{J}^{\rho}\!_{\sigma}\eta_{\rho}
- n \chi^{r}\hat J^{\rho}\!_{r}(g_{\sigma\rho} u +
l^{p}_{\rho}l_{p}u_{\sigma})\; . \label{eq:351}
\end{equation}
On the basis of the evolution equation of the hypermomentum tensor
(\ref{eq:33}) and with the help of (\ref{eq:32}) the expression (\ref{eq:351})
reads,
\begin{equation}
\Sigma_{\sigma} = p\eta_{\sigma} + \frac{1}{c^{2}}(\varepsilon + p) u_
{\sigma} u  + \frac{1}{c^{2}}n g_{\alpha [\sigma}\dot{J}^{\alpha}\!_
{\beta]}u^{\beta}u \; . \label{eq:36}
\end{equation}
After some algebra one can get the other form of the canonical
energy-momentum 3-form,
\begin{equation}
\Sigma_{\sigma} =  p \eta_{\sigma} + \frac{1}{c^{2}}(\varepsilon + p)
u_{\sigma} u + \frac{1}{c^{2}}n \dot{S}_{\sigma}\!_{\beta} u^{\beta} u
- \frac{1}{c^{2}} n J^{\beta}\!_{[\sigma} Q_{\alpha ]\beta \gamma}
u^{\gamma} u^{\alpha} u \; , \label{eq:37}
\end{equation}
where $Q_{\alpha \beta \gamma}$ are components of a nonmetricity 1-form,
\begin{equation}
{\cal Q}_{\alpha \beta } := - {\cal D} g_{\alpha \beta } =
Q_{\alpha \beta \gamma} \theta^{\gamma}\; . \label{eq:nem}
\end{equation}
\par
     The metric stress-energy 4-form can be derived in the same way,
\begin{eqnarray}
&& \sigma^{\alpha\beta} := 2\frac{\delta {\cal L}_{m}}{\delta g_{\alpha\beta}}
= T^{\alpha\beta} \eta \; , \nonumber \\
&& T^{\alpha\beta} = - \varepsilon g^{\alpha\beta} + 2 n \lambda (u^{\alpha}
u^{\beta} + c^{2}g^{\alpha\beta}) - 2 n \chi^{r} \hat J^{(\alpha}\!_{r}
u^{\beta )} \nonumber \\
&& = p g^{\alpha\beta} + \frac{1}{c^{2}}(\varepsilon + p)u^{\alpha}u^{\beta}
+ \frac{1}{c^{2}}n \dot{J}^{(\alpha}\!_{\gamma} u^{\beta )} u^{\gamma} \; .
\label{eq:35}
\end{eqnarray}
\par
     For the hypermomentum 3-form one finds
\begin{equation}
{\cal J}^{\alpha}\!_{\beta} := - \frac{\delta{\cal L}_{m}}{\delta\Gamma^
{\beta}\!_{\alpha}} = \frac{1}{2} n J^{\alpha}\!_{\beta} u \; .
\label{eq:38}
\end{equation}
\par
     Let us consider the special case of the perfect spin fluid with dilatonic
charge, a fluid element of which does not possess the specific shear momentum
tensor, $\hat J^{(p}\!_{q)} = 0$, and is endowed only with the specific spin
momentum tensor $S^{p}\!_{q}$ and the specific dilatonic charge $J$. In this
case the canonical energy-momentum 3-form (\ref{eq:36}) reads,
\begin{equation}
\Sigma_{\sigma} = p\eta_{\sigma} + \frac{1}{c^{2}}(\varepsilon + p) u_
{\sigma} u  + \frac{1}{c^{2}}n g_{\alpha [\sigma}\dot{S}^{\alpha}\!_
{\beta]}u^{\beta}u \; , \label{eq:40}
\end{equation}
where the specific energy density $\varepsilon$ contains the energy density
of the  dilatonic  interaction of the fluid.  This expression coinsides with
the expression of the canonical energy-momentum 3-form of the perfect
dilaton-spin fluid obtained in Ref.\onlinecite{dil}. If the dilatonic charge
also vanishes,  $J  = 0$, then the expression (\ref{eq:40}) will describe
the canonical energy-momentum 3-form of the Weyssenhoff perfect  spin  fluid
in a metric-affine space.

\section{The hydrodynamic equation of motion of the perfect hypermomentum
         fluid}
\markright{Perfect hypermomentum fluid}
     As in case of the perfect dilaton-spin fluid\cite{part} the hydrodynamic
Euler-type equation of motion of the  perfect  hypermomentum  fluid  can  be
derived as the consequence of the covariant energy-momentum quasi-conservation
law. In a general metric-affine space the matter Lagrangian obeys the
diffeomorfism invariance and the local $GL(4,R)$-gauge invariance that leads
(when the equations of matter motion are  fulfilled)  to  the  corresponding
Noether identities (see Ref.\onlinecite{He:pr} and references therein):
\begin{eqnarray}
&&{\cal D}\Sigma_{\sigma} = (\bar{e}_{\sigma}\rfloor {\cal T}^{\alpha})\wedge
\Sigma_{\alpha} - (\bar{e}_{\sigma}\rfloor {\cal R}^{\alpha}\!_{\beta})\wedge
{\cal J}^{\beta}\!_{\alpha} - \frac{1}{2}(\bar{e}_{\sigma}\rfloor {\cal Q}
_{\alpha\beta})\sigma^{\alpha\beta}\; ,  \label{eq:zak1} \\
&&{\cal D}{\cal J}^{\alpha}\!_{\beta} - \theta^{ \alpha}\wedge \Sigma_{\beta}
+ \sigma^{\alpha}\!_{\beta} = 0\; , \label{eq:zak2}
\end{eqnarray}
where ${\cal R}^{\alpha}\!_{\beta}$ is a curvature 2-form and
${\cal T}^{\alpha}$ is a torsion 2-form.
It can be verified that the expressions of the canonical energy-momentum
3-form (\ref{eq:36}), the metric stress-energy 4-form (\ref{eq:35}) and the
hypermomentum 3-form (\ref{eq:38}) are compatible in the sense that they
satisfy to the Noether identities (\ref{eq:zak1}) and (\ref{eq:zak2}).
\par
     Let us  introduce  a  specific  (per  particle) dynamical momentum of a
fluid element,
\begin{eqnarray}
&&\pi_{\sigma}\eta := -\frac{1}{nc^{2}}*\!u\wedge \Sigma_{\sigma}\;, \\
&& \pi_{\sigma} = \frac{\varepsilon}{nc^{2}}u_{\sigma} -
\frac{1}{c^{2}}S_{\sigma\rho}\bar{u}\rfloor {\cal D}u^{\rho} -
\frac{1}{2c^{2}}\hat{J}^{\lambda}\!_{\sigma} u^{\gamma}\bar{u}\rfloor
{\cal Q}_{\lambda\gamma}\; . \label{eq:pi}
\end{eqnarray}
Then the canonical energy-momentum 3-form (\ref{eq:36}) reads,
\begin{equation}
\Sigma_{\sigma} =  p\eta_{\sigma} + n \left (\pi_{\sigma} +
\frac{p}{nc^{2}}u_{\sigma}\right ) u\; . \label{eq:sig}
\end{equation}
\par
     Let us substitute (\ref{eq:sig}), (\ref{eq:35}) and (\ref{eq:38}) into
(\ref{eq:zak1}) and take into account the fluid particles number conservation
law (\ref{eq:7}) and the equality
\begin{equation}
{\cal D}\eta_{\alpha} = {\cal T}^{\beta}\wedge\eta_{\alpha\beta} - \frac{1}{2}
{\cal Q}\wedge \eta_{\alpha}\;,
\end{equation}
where according to Trautman\cite{Tr1} 2-form fields $\eta_{\alpha\beta}$ are
used,
\begin{equation}
\eta_{\alpha\beta} = \bar{e}_{\beta}\rfloor \eta_{\alpha} =
*(\theta_{\alpha}\wedge \theta_{\beta}) \; , \qquad
\theta^{\sigma} \wedge \eta_{\alpha\beta} = -2\delta^{\sigma}_{[\alpha}
\eta_{\beta ]} \;, \label{eq:e2}
\end{equation}
and a Weyl 1-form ${\cal Q}$ is introduced,
\begin{equation}
{\cal Q} := g^{\alpha\beta}{\cal Q}_{\alpha\beta}\;, \qquad
{\cal Q} = Q_{\alpha}\theta^\alpha\;. \label{eq:Q}
\end{equation}
After some algebra one obtains the equation of motion of the perfect
hypermomentum fluid in the form of the generalized hydrodynamic Euler-type
equation,
\begin{eqnarray}
&& u\wedge {\cal D} \left (\pi_{\sigma} + \frac{p}{nc^{2}}u_{\sigma}\right )
= \frac{1}{n}\eta\bar{e}_{\sigma}\rfloor dp
- (\bar{e}_{\sigma}\rfloor{\cal T}^{\alpha}) \wedge \left (\pi_{\alpha}
+ \frac{p}{nc^{2}}u_{\alpha}\right) u \nonumber \\
&& + \frac{1}{2} (\bar{e}_{\sigma}\rfloor {\cal R}^{\alpha}\!_{\beta})\wedge
J^{\beta}\!_{\alpha}u + \frac{1}{2}(\bar{e}_{\sigma}\rfloor
{\cal Q}_{\alpha\beta})\left (\frac{\varepsilon + p}{nc^2}u^{\alpha}u^\beta
+ \frac{1}{c^{2}}(\bar{u}\rfloor{\cal D}J^{\alpha}\!_\gamma ) u^{\beta}
u^{\gamma}\right ) \eta \; . \label{eq:euler}
\end{eqnarray}
\par
     The equation of the hypermomentum fluid motion (\ref{eq:euler}) has the
important consequence,  which can be derived with the help of the procedure
of the decomposition of the connection,\cite{He:pr}
\begin{equation}
\Gamma^{\alpha}\!_{\beta} = \stackrel{C}{\Gamma}\!^{\alpha}\!_{\beta} +
\Delta^{\alpha}\!_{\beta}\; , \qquad
\Delta^{\alpha}\!_{\beta} = g^{\alpha\gamma}\left (\frac{1}{2}
{\cal Q}_{\gamma\beta } - \bar{e}_{[\gamma}\rfloor {\cal Q}_{\beta ]\mu}
\theta^{\mu} \right )\; , \label{eq:def}
\end{equation}
where $\stackrel{C}{\Gamma}\!^{\alpha}\!_{\beta}$   denotes  a  connection
1-form of a Riemann--Cartan space $U_{4}$ with curvature, torsion and
metric compatible with connection, and $\Delta^{\alpha}\!_{\beta}$ is so
called a connection defect 1-form. The decomposition (\ref{eq:def}) of the
connection induces corresponding decomposition of the curvature,
\begin{eqnarray}
&&{\cal R}^{\alpha}\!_{\beta} = \stackrel{C}{{\cal R}}\!^{\alpha}\!_{\beta} +
\stackrel{C}{{\cal D}}\Delta^{\alpha}\!_{\beta}
+ \Delta^{\alpha}\!_{\gamma}\wedge \Delta^{\gamma}\!_{\beta} =
\stackrel{C}{{\cal R}}\!^{\alpha}\!_{\beta} + \frac{1}{4}
\delta^{\alpha}_{\beta}{\cal R}^{\tau}\!_{\tau} + {\cal P}^{\alpha}\!_{\beta}
\;, \quad {\cal P}^{\alpha}\!_{\beta} = {\cal P}^{[\alpha}\!_{\beta ]}\; ,
\label{eq:red1} \\
&&
{\cal R}^{\tau}\!_{\tau} = \frac{1}{2}{\cal D}{\cal Q} =
\frac{1}{2}(\bar{e}_{\lambda}\rfloor {\cal D}Q_{\tau})\theta^{\lambda}\wedge
\theta^{\tau} + \frac{1}{2}Q_{\tau}{\cal T}^{\tau} =
\frac{1}{2}d{\cal Q}\;, \label{eq:seg}
\end{eqnarray}
where $\stackrel{C}{{\cal D}}$ is the exterior covariant differential with
respect to the Riemann--Cartan connection 1-form $\stackrel{C}{\Gamma}\!^
{\alpha}\!_{\beta}$, and $\stackrel{C}{{\cal R}}\!^{\alpha}\!_{\beta}$ is the
Riemann--Cartan curvature 2-form, ${\cal R}^{\tau}\!_{\tau}$ is the Weyl
homothetic curvature 2-form. The corresponding decomposition  of the specific
dynamical momentum of a fluid element (\ref{eq:pi}) reads,
\begin{equation}
\pi_{\sigma} = \frac{\varepsilon}{nc^{2}}u_{\sigma} + \frac{1}{c^{2}}
(\bar{u}\rfloor \stackrel{C}{{\cal D}}S_{\sigma\rho}) u^{\rho}
- \frac{1}{2c^{2}}(\bar{u}\rfloor \Delta^{\rho\lambda})
\hat{J}\!_{\sigma\rho} u_{\lambda} - \frac{1}{2c^{2}} (\bar{u}\rfloor
\Delta^{\lambda\rho}) \hat{J}\!_{\rho\sigma} u_{\lambda}\; . \label{eq:pi1}
\end{equation}
\par
     {\bf Theorem 1.} If the equation of motion of the perfect hypermomentum
fluid (\ref{eq:euler}) is valid (the fluid particles number conservation law
$d (n u) = 0$ being fulfilled), then the energy evolution equation along  a
streamline of the fluid reads,
\begin{equation}
\dot\varepsilon = \frac{\varepsilon + p}{n} \dot{n}\; ,  \label{eq:ev}
\end{equation}
the motion  of  the  perfect  hypermomentum  fluid  is  isoentropic  and the
hypermomentum tensor evolution does not contribute to the energy  change  of
a fluid element. \newline
{\bf Proof}. Let us evaluate the component of the  equation  (\ref{eq:euler})
along the 4-velocity by contracting one with $u^{\sigma}$ and then use the
decompositions (\ref{eq:def}), (\ref{eq:pi1}) and the Frenkel conditions
(\ref{eq:110}), (\ref{eq:111}). The left-hand side of the equation obtained
reads,
\begin{eqnarray}
&& u^\sigma u\wedge {\cal D}\left (\pi_\sigma + \frac{p}{nc^2}\right ) =
(\bar{u}\rfloor d)\left (\frac{\varepsilon + p}{n}\right )\eta
+ (\bar{u}\rfloor \Delta^{(\sigma\rho )})\left (\frac{\varepsilon
+ p}{nc^{2}}\right ) u_\sigma u_{\rho} \eta \nonumber \\
&& - \frac{1}{c^2} (\bar{u}\rfloor\Delta^{(\sigma\rho )}) u_{\sigma}
\hat{J}_{\rho\lambda} (\bar{u}\rfloor \stackrel{C}{{\cal D}}u^\lambda )\eta
- \frac{1}{c^2}(\bar{u}\rfloor\Delta^{(\alpha\sigma)}) u_\sigma (\bar{u}
\rfloor\Delta^{\beta\rho }) u_\rho \hat{J}_{\alpha\beta}\eta\;.\label{eq:ls}
\end{eqnarray}
At the right-hand side  of  the  equation  (\ref{eq:euler})  the  term  with
torsion (the  second one) and the term with curvature (the third one) vanish
after contracting with $u^{\sigma}$. The other terms read,
\begin{eqnarray}
&&\frac{1}{n}\eta (\bar{u}\rfloor dp) + \frac{1}{2}(\bar{u}\rfloor{\cal Q}^
{\sigma\rho })\left (\frac{\varepsilon + p}{nc^2}\right ) u_\sigma u_{\rho}
\eta \nonumber \\
&& - \frac{1}{2c^2} (\bar{u}\rfloor{\cal Q}^{\sigma\rho }) u_\sigma
\hat{J}_{\rho\lambda} (\bar{u}\rfloor \stackrel{C}{{\cal D}}u^\lambda )\eta
- \frac{1}{2c^2}(\bar{u}\rfloor{\cal Q}^{\alpha\sigma}) u_\sigma (\bar{u}
\rfloor\Delta^{\beta\rho }) u_\rho \hat{J}_{\alpha\beta} \eta\;.\label{eq:rs}
\end{eqnarray}
Equating (\ref{eq:ls}) and (\ref{eq:rs}) and taking into account the relation
\begin{equation}
\Delta_{(\sigma\rho)} = \frac{1}{2}{\cal Q}_{\sigma\rho}\;,
\end{equation}
one obtains the equation,
\begin{equation}
\bar{u}\rfloor d\left (\frac{\varepsilon + p}{n}\right ) = \frac{1}{n}
\bar{u}\rfloor d p\;. \label{eq:res}
\end{equation}
As in case of the usual perfect fluid\cite{Miz} this equation means that
along a streamline of the fluid the energy evolution equation takes the form
(\ref{eq:ev}). Comparing the equation (\ref{eq:ev}) with the first
thermodynamic principle (\ref{eq:13}), one can conclude that along a
streamline of the fluid the conditions
\begin{equation}
\dot{s} = 0\;, \qquad
\frac{\partial \varepsilon}{\partial J^{p}\!_{q}} \dot{J}^{p}\!_{q} = 0
\label{eq:comp}
\end{equation}
are valid. The first of these equalities means that the entropy conservation
law is fulfilled along a streamline of the fluid and therefore that the
motion of the perfect hypermomentum fluid is isoentropic. The second of these
equalities means that the hypermomentum tensor evolution does not contribute
to the energy change of the fluid element. As was to be proved.
\par{\bf Remark.}  The  first equation (\ref{eq:comp}) is the consequence of
(\ref{eq:7}). The second equation (\ref{eq:comp})  can  be  derived  as  the
consequence of the fluid motion equation (\ref{eq:23}), the generalized
Frenkel conditions (\ref{eq:110}), (\ref{eq:111}) and the evolution equation
of the hypermomentum tensor (\ref{eq:33}). Therefore the conclusion
(\ref{eq:comp}) of the Theorem 1 means the consistency of the theory.

\section{Peculiarities of the hypermomentum fluid motion}
\markright{Perfect hypermomentum fluid}
\setcounter{equation}{0}
     {\bf Theorem 2.} When the traceless hypermomentum tensor vanishes,
$\hat J^{p}\!_{q} = 0$, the motion of the perfect hypermomentum fluid in a
metric-affine space $(L_{4},g)$ obeys the equation,
\begin{equation}
u \wedge \stackrel{R}{{\cal D}}\left (\frac{\varepsilon + p}{nc^2}
u_\sigma\right ) = \frac{1}{n}\eta \bar{e}_\sigma\rfloor d p
+ \frac{1}{16}(\bar{e}_\sigma \rfloor d {\cal Q}) \wedge J u \;, \label{eq:t2}
\end{equation}
where $\stackrel{R}{{\cal D}}$ is the exterior covariant differential  with
respect to a Riemann (Levi--Civita) connection 1-form
$\stackrel{R}{\Gamma}\!^{\alpha}\!_{\beta}$.\newline
{\bf Proof.} Using  the decomposition the nonmetricity 1-form on the Weyl's
piece and the traceless piece,
\begin{equation}
{\cal Q}_{\alpha\beta} = {\tilde{\cal Q}}_{\alpha\beta} + \frac{1}{4}g_
{\alpha\beta}{\cal Q}\;,\qquad g^{\alpha\beta}{\tilde{\cal Q}}_{\alpha\beta}
=0\;, \label{eq:decq}
\end{equation}
let us  represent the connection 1-form of a metric-affine space $(L_{4},g)$
as follows,
\begin{equation}
\Gamma^{\alpha}\!_{\beta} = \stackrel{C}{\Gamma}\!^{\alpha}\!_{\beta} +
\stackrel{W}{\Delta}\!^{\alpha}\!_{\beta} + \tilde{\Delta}^{\alpha}\!_{\beta}
\; , \qquad \tilde{\Delta}^{\alpha}\!_{\alpha} = 0 \;, \label{eq:def1}
\end{equation}
where the connection defect 1-form corresponding to a connection 1-form of a
Weyl space is introduced,
\begin{equation}
\stackrel{W}{\Delta}\!^{\alpha}\!_{\beta} = \frac{1}{8} (\delta^{\alpha}_
{\beta}{\cal Q} + 2\theta^{[\alpha}Q_{\beta ]}) \;. \label{eq:def2}
\end{equation}
In turn, the Riemann--Cartan connection 1-form can be decomposed as follows,
\begin{eqnarray}
&&\stackrel{C}{\Gamma}\!^{\alpha}\!_{\beta} = \stackrel{R}{\Gamma}\!^
{\alpha}\!_{\beta} + {\cal K}^{\alpha}\!_{\beta}\;, \quad
{\cal T}^{\alpha} =: {\cal K}^{\alpha}\!_{\beta}\wedge \theta^{\beta}\;,
\label{eq:kon1}\\
&& {\cal K}_{\alpha\beta} = 2\bar{e}_{[\alpha}\rfloor {\cal T}_{\beta ]} -
\frac{1}{2} \bar{e}_{\alpha}\rfloor \bar{e}_{\beta}\rfloor ({\cal T}_{\gamma}
\wedge \theta^{\gamma})\;, \label{eq:kon2}
\end{eqnarray}
where ${\cal  K}^{\alpha}\!_{\beta}$ is a kontorsion 1-form.\cite{He:pr}
Taking into account the decompositions (\ref{eq:decq}), (\ref{eq:def1}),
(\ref{eq:kon1}) and the expression for the specific dynamical momentum of a
fluid element (\ref{eq:pi}), the hydrodynamic equation of motion of the
hypermomentum fluid (\ref{eq:euler}) in case of $\hat J^{p}\!_{q} = 0$ reads,
\begin{eqnarray}
&& u\wedge (\delta^\lambda_\sigma \stackrel{R}{\cal D} - K^{\lambda}\!_
{\sigma} - \stackrel{W}{\Delta}\!^{\lambda}\!_{\sigma} - \tilde{\Delta}^
{\lambda}\!_ {\sigma}) \left (\frac{\varepsilon + p}{nc^2} u_\lambda\right )
= \frac{1}{n}\eta \bar{e}_\sigma\rfloor d p + \frac{1}{8}(\bar{e}_\sigma
\rfloor {\cal R}^\lambda\!_\lambda) \wedge J u \nonumber \\
&& - (\bar{e}_\sigma\rfloor {\cal T}^\lambda )\wedge\left (\frac{\varepsilon
+ p}{nc^2}\right ) u_\lambda u - \frac{1}{8}(\bar{e}_\sigma\rfloor {\cal Q})
\left (\frac{\varepsilon + p}{n}\right ) \eta + \frac{1}{2} (\bar{e}_\sigma
\rfloor\;{\tilde{\cal Q}}_{\alpha\beta})\left (\frac{\varepsilon + p}{nc^2}
\right ) u^\alpha u^\beta \eta \;. \label{eq:neu}
\end{eqnarray}
By the direct calculations with the help of (\ref{eq:def}) and
(\ref{eq:kon2}) one can check the equalities,
\begin{eqnarray}
&& u\wedge K^{\lambda}\!_{\sigma}u_\lambda = (\bar{e}_\sigma\rfloor
{\cal T}^\lambda )\wedge u_\lambda u\;, \label{eq:rav1}\\
&& u\wedge \stackrel{W}{\Delta}\!^{\lambda}\!_{\sigma} u_\lambda =
\frac{c^2}{8} (\bar{e}_\sigma\rfloor {\cal Q}) \eta \;, \label{eq:rav2}\\
&& u\wedge \tilde{\Delta}^{\lambda}\!_ {\sigma} u_\lambda = - \frac{1}{2}
(\bar{e}_\sigma\rfloor\;{\tilde{\cal Q}}_{\alpha\beta}) u^\alpha u^\beta \eta
\;. \label{eq:rav3}
\end{eqnarray}
On the basis of the equalities (\ref{eq:rav1})--(\ref{eq:rav3}) and
(\ref{eq:seg}) the equation (\ref{eq:neu}) yields the equation (\ref{eq:t2}),
as was to be proved.
\newline
{\bf Corollary 1.} When the traceless hypermomentum tensor vanishes, the
motion of the perfect hypermomentum fluid in a metric-affine space $(L_{4},g)$
coincides with the motion of the perfect fluid with dilatonic charge in
a Weyl--Cartan space $Y_4$, the metric tensor, the torsion 2-form and the
Weyl 1-form of which coinside with the metric tensor, the torsion 2-form and
the Weyl 1-form of $(L_{4},g)$, respectively.\newline
{\bf Corollary  2.} When the traceless hypermomentum tensor vanishes, the
motion of the perfect hypermomentum fluid in a metric-affine space $(L_{4},g)$
coincides with the motion of the perfect fluid with dilatonic charge in
a Weyl space $W_4$, the metric tensor and the Weyl 1-form of which coinside
with the metric tensor and the Weyl 1-form of $(L_{4},g)$, respectively.
\newline
{\bf Proof.} These two statments are the consequences of the fact that in the
equalities (\ref{eq:rav1})--(\ref{eq:rav3}) torsion and traceless
nonmetricity are arbitrary and can be put to be equal to zero. If only the
traceless 1-form is equal to zero one gets a Weyl--Cartan space. If both the
nonmetricity traceless 1-form and the torsion 2-form are  equal to zero,
then one gets a Weyl space. In these cases the equation of motion of the
perfect hypermomentum fluid (\ref{eq:t2}) in a metric-affine space $(L_{4},g)$
coincides with the equations of motion of the perfect fluid with dilatonic
charge in $Y_4$ and $W_4$,  respectively,  which are the particular cases
(when $S_{\alpha\beta} = 0$) of the equation of motion of the perfect
dilaton-spin fluid derived in Ref.\onlinecite{part}.
\newline
{\bf Corollary 3.} If all irreducible pieces of the hypermomentum tensor
vanish, $J_{\alpha\beta}  = 0$,  then the equation of motion of the perfect
hypermomentum fluid in a metric-affine space $(L_{4},g)$ coincides with
the equation of motion of the usual perfect fluid in a Riemann space $V_4$,
the metric tensor of which coinsides with the metric tensor of $(L_{4},g)$.
\newline
{\bf Proof.} If $J = 0$ the equation (\ref{eq:t2})  with  the  help  of  the
Theorem 1 yields the equation of motion of the usual perfect fluid in a
Riemann space,\cite{Miz}
\begin{equation}
\frac{1}{c^2}(\varepsilon + p)\bar{u}\rfloor \stackrel{R}{{\cal D}}u_\sigma =
- \Pi^\rho_\sigma \bar{e}_\rho\rfloor d p\;, \label{eq:t3}
\end{equation}
where $\Pi^{\rho}_{\sigma}$ is the projection tensor (\ref{eq:121}).
\newline
{\bf Corollary 4.} The motion of the test particle with dilatonic charge  in
a metric-affine space $(L_{4},g)$ coincides with the motion of this particle
both in a Weyl--Cartan space $Y_4$ and in a Weyl space $W_4$; if the dilatonic
charge of the particle vanishes, then the motion of this test particle in
$(L_{4},g)$ is realized along geodesics of a Riemann space $V_4$, the metric
tensor of which coincides with the metric tensor of $(L_{4},g)$.
\newline
{\bf Proof.} The equation of motion of the test particle with dilatonic
charge and with the mass $m_{0} = \varepsilon/(n c^{2}) = const$ in a
metric-affine space is obtained as the limiting case of the equation of
motion (\ref{eq:t2}) when the pressure $p$ vanishes. The statement of the
Corollary 4 follows from the statements of the Corollaries 1--3.
\par
     The statements of the Corollaries 3 and 4 are the particular cases of
the Theorem  stated in Refs.\onlinecite{Lan} --\onlinecite{BKF2}  for the
matter motion in a general metric-affine space.
\par
     {\bf Theorem.} In $(L_{4},g)$ the motion of matter without
hypermomentum coinsides with the motion in a Riemann space-time $V_4$,
the metric tensor of which coinsides with the metric tensor of $(L_{4},g)$.

\section{Conclusions}
\markright{Perfect hypermomentum fluid}
\setcounter{equation}{0}
     The essential  feature  of  the  constructed  variational theory of the
hypermomentum perfect fluid is the assumption that the frame realized by all
four  directors  is  elastic.  The  deformation  of the directors during the
motion of the  fluid  element,  from  one  side,  generates  the  space-time
nonmetricity and, from the other side, allows nonmetricity of the space-time
to be discovered.  As the consequence of this fact  the  Lagrangian  density
(\ref{eq:20}) does not contain the term maintaining the orthogonality of the
directors.  The  time-like  director  needs  not  to  be  collinear  to  the
4-velocity  of  the fluid element.  The essential feature of our variational
approach is the using the Frenkel conditions for the traceless hypermomentum
tensor in the form (\ref{eq:110}) and (\ref{eq:111}), which do not coincide
nor with their classical form, when the Frenkel condition is imposed on the
spin tensor, $S_{\alpha\beta} u^{\beta}=0$, nor with its generalized
form, when the Frenkel condition is imposed on the full intrinsic
hypermomentum tensor, $J_{\alpha\beta} u^{\beta} = J_{\alpha\beta}
u^{\alpha} =~0$.
\par
     The expression of the energy-momentum tensor of the fluid (\ref{eq:36})
coincides with one obtained in the other approaches.\cite{Ob-Tr,Bab:tr3} But
our approach does not contain the shortcomings,  which are inherent  in  the
previous  theories  of  the hypermomentum perfect fluid.  First of all,  our
variational theory contains the Weyssenhoff spin  fluid  as  the  particular
case  that  is  important from physical point of view.  Then,  the evolution
equation  of  the  hypermomentum  tensor  (\ref{eq:33})   demonstrates   the
Weyssenhoff-type dynamics.  At last,  the perfect hypermomentum fluid theory
developed allows to describe as the special case the perfect spin fluid with
dilatonic charge.  It should be important to investigate the consequences of
the employing the perfect fluid of such  type  as  the  gravitational  field
source in cosmological and astrophysical problems.
\par
    The perfect hypermomentum fluid model represents the medium with spin,
shear momentum and dilatonic charge which  generate the  space-time
metric-affine geometrical structure and interact with it. The influence of
this geometry on fluid motion is described by the Euler-type hydrodynamic
equation (\ref{eq:euler}), the properties of which are established by
Theorem 1 of Sec. 6. The main features of this motion are its isoentropic
character and the fact that the hypermomentum tensor evolution does not
contribute to the energy change of the fluid element.
\par
     The peculiarities  of  the hypermomentum fluid motion are discovered in
Theorem 2 of Sec. 7. The important consequences of this Theorem mean that
bodies and mediums without hypermomentum are  not  subjected to the influence
of the possible nonmetricity of the space-time (in contrast to the generally
accepted opinion) and therefore can not be the tools for the detection of the
deviation of the space-time properties from the Riemann space structure.
\par
     Therefore for the investigation of the different manifestations of the
possible space-time nonmetricity one needs to use the bodies and  mediums
endowed with the hypermomentum, i.e. particles and fluids with spin,
dilatonic charge or with intrinsic hypermomentum.


\vskip 0.2cm
\begin{thebibliography}{60}
\bibitem{WR}
J. Weyssenhoff and A. Raabe, {\it Acta Phys. Polon.} {\bf 9}, 7 (1947).
\bibitem{Hal}
E. Halbwachs, {\it Th\`{e}orie relativiste des fluides a spin}
(Gauthier-Villars, Paris, 1960).
\bibitem{Tr}
A. Trautman, {\it Nature} {\bf 242}, 7 (1973).
\bibitem{Kop1}
W. Kopczy\'{n}ski, {\it Phys. Lett.} {\bf A39}, 219 (1972); {\bf A43}, 63
(1973).
\bibitem{Sm}
L.L. Smalley, {\it Phys. Rev.} {\bf D32}, 3124 (1985).
\bibitem{Dem}
M. Demianski, R. de Ritis, G. Platania, P. Scudellaro and C. Stornaiolo,
{\it Phys. Lett.} {\bf A116}, 13 (1983).
\bibitem{Gas}
M. Gasperini, {\it Phys. Rev. Lett.} {\bf 56}, 2873 (1986).
\bibitem{Tun1}
V. N. Tunyak, {\it Dokl. Acad. Nauk BSSR} {\bf 19}, 559 (1975) (in Russian).
\bibitem{Tun2}
V. N. Tunyak, {\it Izvestija vyssh. uch. zaved. (Fizika)}, {\bf N12}, 11
(1977) (in Russian).
\bibitem{Ray-Sm}
J. R. Ray and L. L. Smally, {\it Phys. Rev.} {\bf D27}, 1383 (1983).
\bibitem{Rit}
R. de Ritis, M. Lavorgna, G. Platania, and C. Stornaiolo, {\it Phys. Rev.}
{\bf D28}, 713 (1983).
\bibitem{Ob-Kor}
Yu. N. Obukhov and V. A. Korotky, {\it Class. Quantum Grav.} {\bf 4}, 1633
(1987).
\bibitem{Bab-Fr}
O. V. Babourova, B. N. Frolov, in {\it Gravitatziya i elektromagnetizm}
(Minsk, ``Uni\-ver\-si\-tetskoe'', 1987), p. 3 (in Russian).
\bibitem{Bab:iz}
O. V. Babourova, {\it Izvestija vyssh. uch. zaved. (Fizika)}, {\bf N10},
101 (1989) (in Russian).
\bibitem{Bab:Boul}
O. V. Babourova, B. N. Frolov, in {\it Abstr. cont. papers, 12th Intern.
Conf. on Gen. Rel. and Grav.} (USA, Boulder, 1989), p. 151 (A3:08).
\bibitem{Bab:thes}
O. V. Babourova, ``Variational theory of a perfect fluid with inrinsic
degrees of freedom in modern gravitational theory'', PhD thesis, VNICPV,
Moscow, 1989 (in Russian).
\bibitem{Ol}
H. P. de Oliveira, {\it Gen. Rel. Grav.} {\bf 25}, 473 (1993).
\bibitem{Min-Kar}
A. V. Minkevich and P. Karakura, {\it J. Math. A: Math. Gen.} {\bf 16}, 1409
(1983).
\bibitem{Kop}
W. Kopczy\'{n}ski, {\it Phys. Rev.} {\bf D34}, 352 (1986).
\bibitem{GHK}
J. Gibbons, D.D. Holm and B.A. Kupershmidt, {\it Phys. Lett.} {\bf 90A}, 281
(1982).
\bibitem{Am}
R. Amorim, {\it Phys. Rev.} {\bf D33}, 2796 (1986).
\bibitem{prep}
V.G. Bagrov,  O.V.  Babourova,  A.S.  Vshivtsev, B.N. Frolov, ``Motion of
colour spin  particle  in  non-Abelian  gauge  fields   in   Riemann--Cartan
space-time'', Siberian  Department  of  Academy  of  Sciences of USSR (Tomsk
filiation), Preprint No 33, 1988 (in Russian) (unpublished).
\bibitem{Ind}
O.V. Babourova and B.N. Frolov, {\it Mahavisva} (India) {\bf 4}, N1,2, 55
(1991).
\bibitem{Choq}
Y. Choquet-Bruhat, {\it C. R. Acad. Sci.} (Paris) {\bf 314}, S\'{e}r. I,
87 (1992).
\bibitem{DAN}
O.V. Babourova, A.S. Vshivtsev, V.P. Myasnikov, B.N. Frolov, {\it Doklady
Akademii Nauk} {\bf 357}, No 2, 183 (1997) ({\it Physics--Doklady} {\bf 42}
No 11, 611 (1997)).
\bibitem{Yaf}
O.V. Babourova, A.S. Vshivtsev, V.P. Myasnikov, B.N. Frolov, {\it Yad. Fiz.}
{\bf 61}, No 5, 888 (1998) (in Russian).
\bibitem{VP}
A.S. Vshivtsev, D.V. Peregudov, {\it Yad. Fiz.} {\bf 60}, N7, 1481 (1997)
(in Russian).
\bibitem{dil}
O.V. Babourova and B.N. Frolov, {\it Mod. Phys. Letters A} {\bf 12}, No 38,
2943 (1997) (LANL e-archive gr-qc/9708006).
\bibitem{part}
O.V. Babourova and B.N. Frolov, {\it Mod. Phys. Letters A} {\bf 13}, No 1,
7 (1998) (LANL e-archive gr-qc/9708009).
\bibitem{GSW}
M.B. Green, J.H. Schwarz, E. Witten, {\it Superstring Theory}
(Cambridge Univ. Press, 1987).
\bibitem{He-Ker}
F.W. Hehl,  G.D. Kerlick, P. Heyde, {\it Z. Naturforsch} {\bf 31A} 111, 524,
823 (1976).
\bibitem{Bab1}
O. V. Babourova, in {\it Gravitaciya i Fundamental'nye vzaimodejstviya}
(UDN, Moscow, 1988), p. 119 (in Russian).
\bibitem{Bab:Arg}
O. V. Babourova, B. N. Frolov, M. Yu. Koroliov, in {\it 13th
Int. Conf. gen. rel. grav.}, Abstracts of contributed papers, edited  by
P. W. Lamberti and O. E. Ortiz (Cordoba, Argentina, 1992),  p. 131.
\bibitem{Bab:tr1}
O. V. Babourova, B. N. Frolov, M. Yu. Koroliov, in {\it Materialy Nauchnoj
sessii ... MPGU  za 1991. Ser. Est. nauki} (Moscow, ``Prometey'', 1992), p.
4. (in Russian).
\bibitem{Bab:tr2}
O. V. Babourova, B. N. Frolov, M. Yu. Koroliov, in {\it Nauchnye trudy Mosk.
Ped. Gos. Univ. Ser. Est. nauki} (Moscow, ``Prometey'', 1993), p. 170
(in Russian).
\bibitem{Ob-Tr}
Y. N. Obukhov and R. Tresguerres, Phys. Let. A {\bf 184}, 17 (1993).
\bibitem{Bab:tr3}
O. V. Babourova, B. N. Frolov, M. Yu. Koroliov, in {\it Nauchnye trudy MPGU.
Ser. Est. nauki} (Moscow, ``Prometey'', 1994), Part I, p. 89 (in Russian).
\bibitem{Sm-Kr}
L. L. Smally and J. P. Krisch, J. Math. Phys. {\bf 36}, 778 (1995).
\bibitem{GR14}
O. V. Babourova, B. N. Frolov, in {\it 14th International Conf. on Gen. Rel.
and Grav.}, Abstracts of Contributed Papers (Florence, Italy, 1995), p. A90.
\bibitem{BF:Los}
O. V. Babourova and B. N. Frolov, ``The variational theory of perfect fluid
with intrinsic hypermomentum in space-time with nonmetricity'' (LANL
e-archive gr-qc/9509013, 1995).
\bibitem{Ob2}
Yu. N. Obukhov, Phys. Lett. {\bf A210}, 163 (1996).
\bibitem{Nov}
O. V. Babourova, B. N. Frolov, in {\it Teoreticheskie i eksperimental\'{}nye
problemy gravitacii}, Abstracts of Contr. Pap. 9 Russ.
Grav. Conf., Novgorod (Moscow, 1996), Part 1, p. 44 (in Russian).
\bibitem{Los:pr}
O. V. Babourova and B. N. Frolov, ``Variational theory of perfect
hypermomentum fluid'' (LANL e-archive gr-qc/9612055, 1996).
\bibitem{He:pr}
F. W. Hehl, J. L. McCrea, E. W. Mielke and Yu. Ne\'{ }eman, Phys. Reports
{\bf 258}, 1 (1995).
\bibitem{Ne:Si}
Y. Ne\'{ }eman, Dj. \u{S}ija\u{c}ki, {\it Ann. Phys.} {\bf 120}, 292 (1979).
\bibitem{LN}
C.-Y. Lee and Y. Ne\'{ }eman, {\it Phys. Lett.} {\bf B242}, 59 (1990).
\bibitem{Tr1}
A. Trautman, {\it Symp. Math.} {\bf 12}, 139 (1973).
\bibitem{Tr2}
A. Trautman, {\it Bul. Acad. Pol. Sci. Ser. sci. math., astr., phys.}
{\bf 20}, 895 (1972).
\bibitem{Miz}
C.W. Misner, K.S. Thorne, J.A. Wheeler, {\it Gravitation}
(W.H. Freeman and Company, San Francisco, 1973), V. 2.
\bibitem{Lan}
O.V. Babourova, in: {\it Abstracts of contrib. papers of the Cornelius
Lanczos Int. conf.} (NC State Univ., USA, 1993) p. 100.
\bibitem{BKF1}
O.V.Babourova, M. Yu. Koroliov and B.N. Frolov, {\it Izvestija vyssh. uch.
zaved. (Fizika)} {\bf N1}, 76 (1994) (in Russian).
\bibitem{BKF2}
O. V.  Babourova, B. N. Frolov and M.Yu. Koroliov, ``Peculiarities of matter
motion in metric-affine gravitational theory'' (LANL e-archive
gr-qc/9502012, 1995).
\end{thebibliography}
\end{document}